\documentclass[aps,preprint,prb,floatfix,superscriptaddress]{revtex4-2}
\usepackage[english]{babel}
\usepackage[utf8]{inputenc}
\usepackage[T1]{fontenc}
\usepackage{amssymb}
\usepackage{amsmath}
\usepackage{amsbsy}
\usepackage{bm}
\usepackage{graphics}
\usepackage{graphicx}
\usepackage{braket}
\usepackage{color,soul}
\usepackage{ulem}
\usepackage{wasysym}
\usepackage{mathrsfs}
\usepackage{times}
\usepackage[dvipsnames,svgnames,table]{xcolor}
\usepackage{hyperref}
\usepackage{placeins}

\hypersetup{
pdfstartview={FitH},
colorlinks=true, 
linkcolor=Maroon, 
citecolor=Maroon,
filecolor=Maroon, 
urlcolor=Maroon 
}

\makeatletter
\def\bbl@set@language#1{%
\edef\languagename{%
\ifnum\escapechar=\expandafter`\string#1\@empty
\else\string#1\@empty\fi}%
\@ifundefined{babel@language@alias@\languagename}{}{%
\edef\languagename{\@nameuse{babel@language@alias@\languagename}}%
}%
\select@language{\languagename}%
\expandafter\ifx\csname date\languagename\endcsname\relax\else
\if@filesw
\protected@write\@auxout{}{\string\select@language{\languagename}}%
\bbl@for\bbl@tempa\BabelContentsFiles{%
 \addtocontents{\bbl@tempa}{\xstring\select@language{\languagename}}}%
\bbl@usehooks{write}{}%
\fi
\fi}
\newcommand{\DeclareLanguageAlias}[2]{%
\global\@namedef{babel@language@alias@#1}{#2}%
}

\makeatother
\DeclareLanguageAlias{en}{english}

\begin{document}
\title{Beyond-adiabatic flat Chern bands from a double-helix skyrmion crystal}

\author{Bin Xi}
\thanks{These authors contributed equally to this work}
\affiliation{College of Physics Science and Technology, Yangzhou University, Yangzhou 225002, China}

\author{Ken Chen}
\thanks{These authors contributed equally to this work}
\affiliation{Lanzhou Center for Theoretical Physics, Lanzhou University, Lanzhou 730000, China}
\affiliation{Key Laboratory of Quantum Theory and Applications of MoE, Lanzhou University, Lanzhou 730000, China}
\affiliation{Key Laboratory of Theoretical Physics of Gansu Province $\&$ Gansu Provincial Research Center for Basic Disciplines of Quantum Physics, Lanzhou University, Lanzhou 730000, China}
\affiliation{School of Mathematics and Physics, Southwest University of Science and Technology, Mianyang 621010, China}

\author{Qiang Luo}
\affiliation{College of Physics, Nanjing University of Aeronautics and Astronautics, Nanjing, 211106, China}

\author{Jie Lu}
\affiliation{College of Physics Science and Technology, Yangzhou University, Yangzhou 225002, China}

\author{Jia-Wei Mei}
\affiliation{Department of Physics, Southern University of Science and Technology, Shenzhen 518055, China}

\author{Hong-Gang Luo}
\affiliation{Lanzhou Center for Theoretical Physics, Lanzhou University, Lanzhou 730000, China}
\affiliation{Key Laboratory of Quantum Theory and Applications of MoE, Lanzhou University, Lanzhou 730000, China}
\affiliation{Key Laboratory of Theoretical Physics of Gansu Province $\&$ Gansu Provincial Research Center for Basic Disciplines of Quantum Physics, Lanzhou University, Lanzhou 730000, China}

\author{Jize Zhao}
\email[Corresponding author: ]{zhaojz@lzu.edu.cn}
\affiliation{Lanzhou Center for Theoretical Physics, Lanzhou University, Lanzhou 730000, China}
\affiliation{Key Laboratory of Quantum Theory and Applications of MoE, Lanzhou University, Lanzhou 730000, China}
\affiliation{Key Laboratory of Theoretical Physics of Gansu Province $\&$ Gansu Provincial Research Center for Basic Disciplines of Quantum Physics, Lanzhou University, Lanzhou 730000, China}

\begin{abstract}
A central challenge in flat-band engineering is suppressing kinetic energy without sacrificing Berry curvature. We show that a double-helix skyrmion crystal (DHSKX)---two sublattice-resolved skyrmion textures locked at opposite helicities, obtained here as the classical ground state of a frustrated honeycomb spin model---provides such a route under double exchange. The key mechanism is a single real-space organization, \textit{phase clustering}: the $\pi$-locked helicities expel the wave function's phase winding from the skyrmion cores, and the magnetic $C_3$ symmetry pins it into three phase-locked clusters whose distributed destructive interference cancels net transport while preserving the Berry curvature. Ordinary skyrmion crystals, even with the same symmetry, do not develop this organization. Phase clustering yields isolated flat $|C|=1$ Chern bands over broad coupling windows, one of which surpasses the adiabatic reference in quantum geometry at intermediate coupling. In this beyond-adiabatic window, band-projected exact diagonalization gives finite-size evidence consistent with $\nu=1/3$ Laughlin-type fractional-Chern-insulator physics; the same texture also hosts a higher-Chern ($C=-2$) flat band. Built from site-resolved complex hoppings alone, the DHSKX architecture is directly programmable in topolectric, acoustic, and photonic platforms.
\end{abstract}
\maketitle
\textit{Introduction.---}
Flat Chern bands are a natural starting point for correlation-driven topological phases, including fractional Chern insulators~(FCIs)~\cite{TangMeiWen2011, PhysRevLett.106.236803, Neupert2011, Sheng2011, PhysRevX.1.021014, WangGuGongSheng2011}, but designing them remains subtle because one must suppress kinetic energy without eliminating Berry curvature. Most established routes address this challenge in momentum space, as in moir\'e and superlattice systems~\cite{Yankowitz2019,Bistritzer2011,Andrei2020Dec,Tang2014Dec,Balents2020Jul,li2021,Xie2020Mar,Zhang2019Nov,Wu2019Feb,Yu2020Jan,Zhang2024May,Ghorashi2023}. A related route uses noncoplanar magnetic textures: continuum skyrmion-proximity models can generate flat Chern minibands for both Schr\"odinger and Dirac electrons~\cite{Paul2023Feb,Guan2023,Wang2024Aug}, single skyrmion crystals on triangular lattices already produce Chern bands under double exchange~\cite{Hamamoto2015,Gobel2017}, and engineered bilayer skyrmion crystals can bring $|C|=1$ bands close to the Landau-level limit~\cite{Hardy2025}. These works establish skyrmion textures as an important flat-band platform, while leaving open how a fully discrete, site-resolved magnetic architecture can organize flat Chern bands beyond the adiabatic smooth-texture picture through the real-space structure of the wave function itself. This motivates a search for flat Chern bands whose bandwidth suppression is set by the internal architecture of the magnetic unit cell.

Here we show that such a route is realized by a DHSKX on the honeycomb lattice over an extended range of electron-spin coupling, without relying on a single optimized point. The DHSKX is not an ad hoc imposed texture but emerges as a distinct noncoplanar phase of a frustrated honeycomb $J\Gamma'$ spin model with easy-plane anisotropy. It consists of two sublattice-resolved skyrmion textures with opposite helicities locked into a composite crystal with total topological charge $-2$, making it distinct from both conventional biskyrmions~\cite{Yu2014Biskyrmion} and antiferromagnetic skyrmions~\cite{Zhang2016AFMSk}. Under double exchange, this architecture generates isolated flat Chern bands over extended intervals of $t/J_{\rm H}$, with the same phase-clustered organization persisting in a reduced-period analogue. The central result is therefore a discrete double-helix organizing principle for the recurring $|C|=1$ branches: the $\pi$-locked opposite helicities force their wave functions into $C_3$-pinned phase clusters, and the clustered winding cancels net transport by distributed destructive interference, preserving the complex hopping phases responsible for Berry curvature. Quantum geometry sharpens this picture: one of the recurring clustered $|C|=1$ branches reaches its most Landau-level-like geometry at intermediate coupling, surpassing the adiabatic strong-coupling reference. Two consequences follow. In this geometry-optimal window, band-projected exact diagonalization on accessible tori up to $N_\phi=30$ gives finite-size evidence consistent with $\nu=1/3$ Laughlin-type FCI physics---and the Laughlin signatures open only on the geometry-optimized branch, not on the flattest one; separately, the same texture hosts a higher-Chern $C=-2$ flat band. Moreover, because the DHSKX background maps onto a tight-binding network with site-resolved complex hoppings, the architecture is directly reproducible in topolectric, acoustic, and photonic platforms without requiring a self-organized magnetic state.

\begin{figure}[tbp]
\centering
\includegraphics[width=\columnwidth]{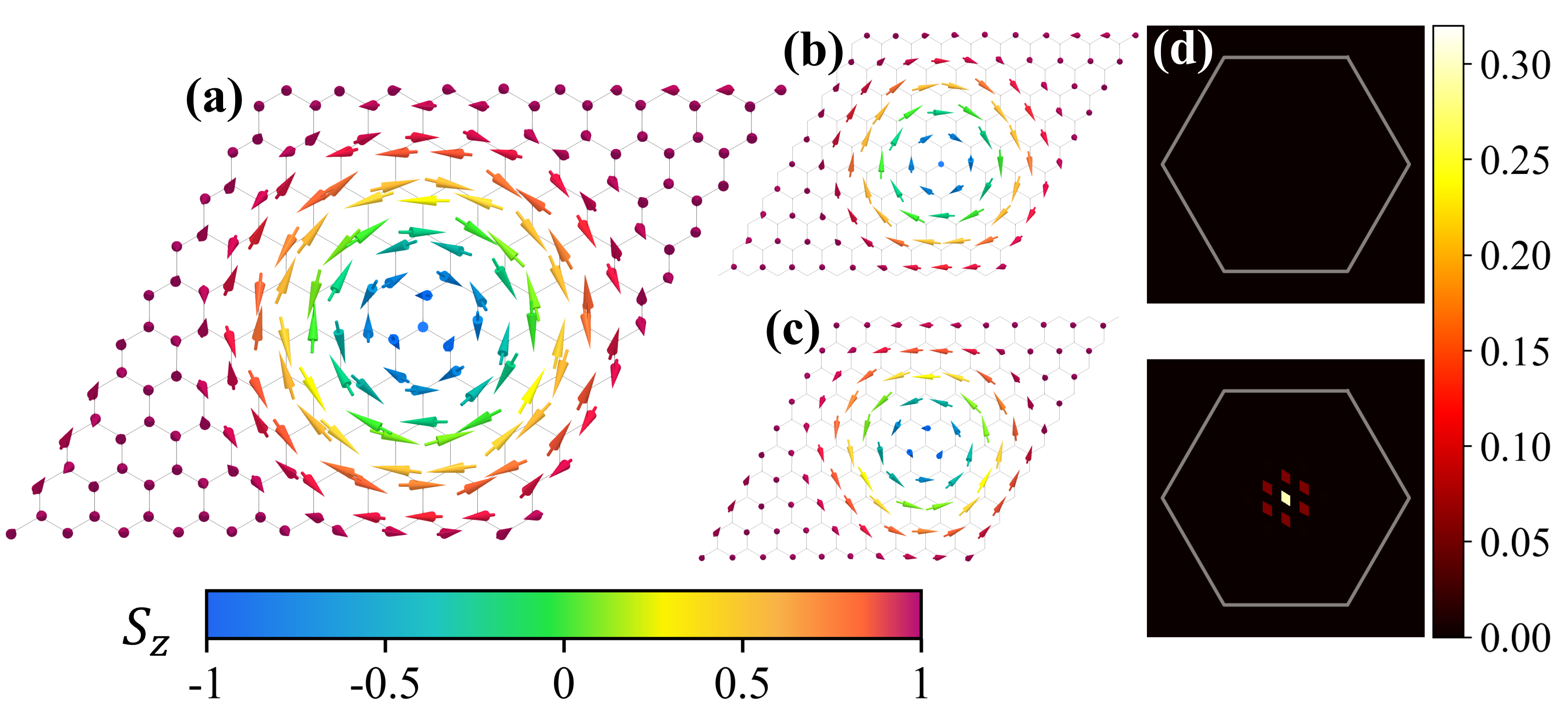}
\caption{
Representative DHSKX configuration. (a) Full spin texture. (b,c) Sublattice-resolved spin textures. (d) Corresponding spin structure factors in the $ab$ plane (upper) and along the $c$ axis (lower).
}
\label{fig1}
\end{figure}

\textit{DHSKX.---} We use a frustrated honeycomb $J\Gamma'$ model with easy-plane anisotropy as a representative microscopic generator of the DHSKX phase. The easy spin plane is the $ab$ plane and the $c$ axis is the perpendicular hard axis. We use a representative parameter point in the stable DHSKX regime.
In the present model, a $c$-axis magnetic field stabilizes a noncoplanar spin configuration on the honeycomb lattice. As shown in Fig.~\ref{fig1}(a), the resulting spin texture contains two entangled skyrmions with opposite helicities residing on the two separate sublattices. The individual sublattice textures [Figs.~\ref{fig1}(b) and \ref{fig1}(c)] reveal Bloch-type skyrmions whose helicities differ by $\pi$, so we refer to the composite phase as a DHSKX. Unlike the Bloch-type skyrmions familiar from Dzyaloshinskii-Moriya systems~\cite{NagaosaTokura2013}, however, the present skyrmion crystal is generated by bond-dependent off-diagonal exchange and easy-plane anisotropy in a frustrated honeycomb $J\Gamma'$ model rather than by chiral spin-orbit coupling alone~\cite{Luo2022May}. A lattice solid-angle evaluation gives skyrmion number (vorticity) approximately $Q_v = -1.0$ on each sublattice, and hence a noncompensated total charge $Q_v^{\rm tot} = -2$.
The spin structure factors (SSFs), defined as ${{\mathcal{F}}_{\bf{q}}^{\mu}} = \frac{1}{N^2} |\sum_{i}e^{i{\bf{q}}\cdot\textbf{r}_i}{S_i^{{\mu}} } |^2$, are shown in Fig.~\ref{fig1}(d).
They differ significantly from those of a conventional skyrmion crystal. In the $ab$-plane (upper panel),
the SSFs $\mathcal{F}_{\mathbf{q}}^{a/b}$ are negligible due to the cancellation from opposite helicities.
In contrast, along the $c$-axis (lower panel), $\mathcal{F}_{\mathbf{q}}^{c}$ exhibits three pairs of Bragg peaks and a $\Gamma$-point peak, indicating a triple-$Q$ state with a ferromagnetic background. The DHSKX evolves from a zero-field coupled spiral state with opposite sublattice chiralities.

\textit{Flat Chern bands from the DHSKX architecture.---} When itinerant electrons are coupled to a spin texture, the Hamiltonian is given by the double-exchange model~\cite{AndersonHasegawa1955}:
\begin{eqnarray}\label{eq5}
	\mathcal{H}_{ex} &=& t\sum_{\langle{i,j}\rangle}\left(c^\dagger_{i\sigma} c_{j\sigma} + \mathrm{h.c.}\right) + J_{\rm{H}}\sum_{i\sigma\sigma'}\mathbf{S}_{i}\cdot c^\dagger_{i\sigma} \boldsymbol{\sigma} c_{i\sigma'},
	\label{HDE} 	
\end{eqnarray}
where $c^\dagger_{i\sigma}$ ($c_{i\sigma}$) is the creation (annihilation) operator of itinerant electrons with spin $\sigma$ at site $i$.
$t$ is the hopping integral and $\langle{i,j}\rangle$ means that the summation is over the nearest-neighbor sites $i$ and $j$.
$J_{\rm{H}}$ is the Hund's coupling between the electron and the on-site localized spin $\mathbf{S}_i$.
Throughout this work we set $t=1$ as the energy unit and vary $J_{\rm H}$, equivalently parameterizing the sweep by $t/J_{\rm H}$.

\begin{figure*}[tbp]
        \centering
        \includegraphics[width=0.97\textwidth]{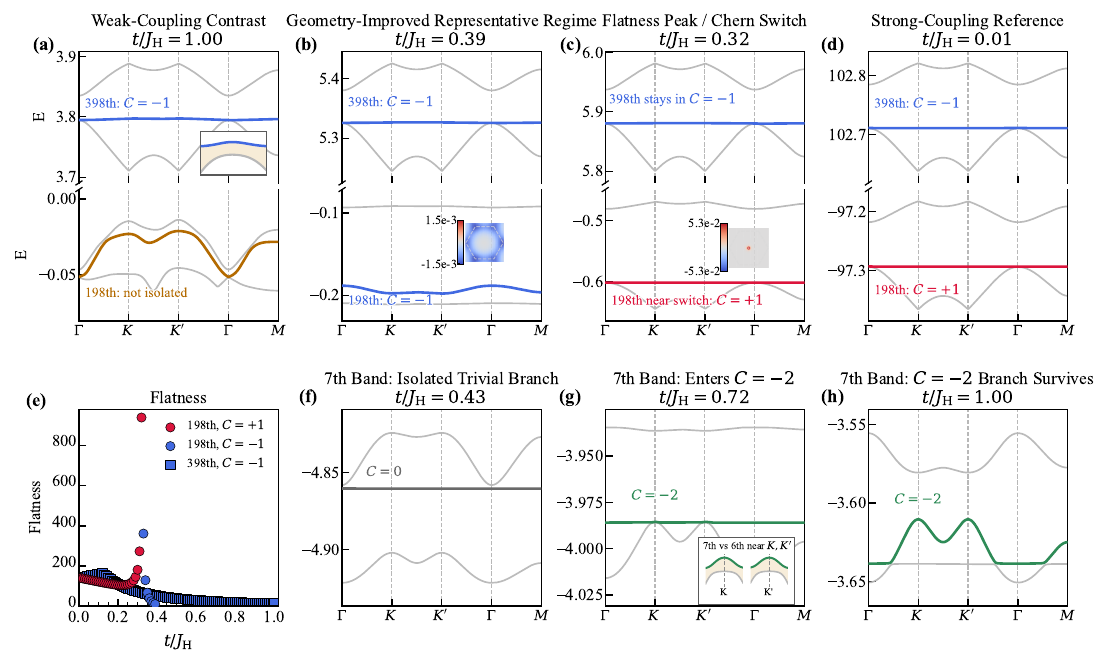}
        \caption{Flat Chern bands in the DHSKX background. (a--d) Narrow energy windows around the 398th (upper) and 198th (lower) bands at $t/J_{\rm H}=1.00$, $0.39$, $0.32$, and $0.01$; the highlighted bands are annotated by their Chern sectors when isolated. Inset in (a): enlarged lower direct gap of the 398th band near $\Gamma$ on the return path. Insets in (b) and (c): Berry-curvature maps of the 198th band at $t/J_{\rm H}=0.39$ and $0.32$, respectively. (e) Flatness of the 198th and 398th branches versus $t/J_{\rm H}$. (f--h) Conversion of the 7th band from a perfectly flat trivial band at $t/J_{\rm H}=0.43$ to a $C=-2$ flat band at $t/J_{\rm H}=0.72$ and $1.00$. Inset in (g): enlarged direct gaps between the 7th and 6th bands near $K$ and $K'$.}
        \label{fig2}
\end{figure*}
\begin{figure}[tbp]
        \centering
        \includegraphics[width=\columnwidth]{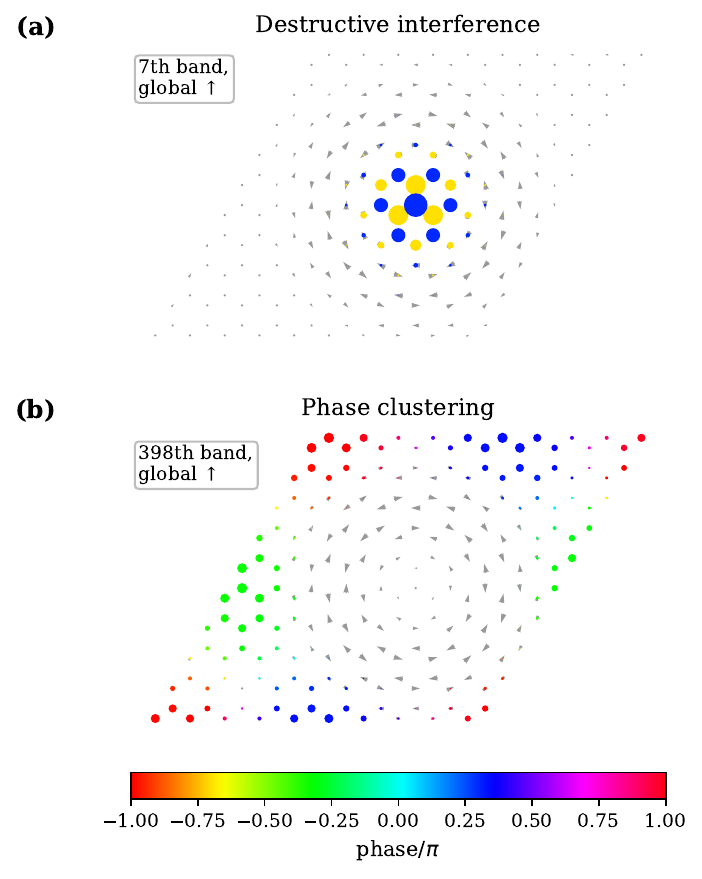}
        \caption{Real-space wave-function fingerprints relevant to Fig.~\ref{fig2}. (a) Global $\uparrow$ component of the 7th band at $t/J_{\rm H}=0.43$, showing a core-localized antiphase pattern. (b) Global $\uparrow$ component of the isolated 398th band at $t/J_{\rm H}=0.01$, showing extended regions of nearly uniform phase with relative phase offsets close to $\pm2\pi/3$. For both bands the wave-function weight lies almost entirely in the global $\uparrow$ channel, which is the one shown here. Marker size denotes wave-function amplitude, color denotes phase, and gray arrows show the underlying spin texture.}
        \label{fig3}
\end{figure}

The DHSKX configuration shown in Fig.~\ref{fig1}(a) serves as the representative example throughout. Its unit cell contains $10 \times 10 \times 2$ spins,
giving rise to a total of 400 electronic bands. We determine the band Chern numbers by combining the zero-temperature Kubo Hall response~\cite{Kubo1957Jun} with the single-band Fukui-Hatsugai-Suzuki construction~\cite{Fukui2005}. We use these two diagnostics to identify isolated Chern sectors throughout the coupling scan. 

The central observation is that this fixed DHSKX background produces isolated flat Chern bands over extended intervals of $t/J_{\rm H}$. For nonzero $J_{\rm{H}}$, the 400 bands separate into two distinct blocks with an energy gap in between. Electron spins in the lower block tend to align antiparallel to the local magnetic texture, while those in the upper block prefer parallel alignment.
Near the top of each block, there is one band---the 198th in the lower block and the 398th in the upper, hereafter the lower and upper \textit{block-edge} bands---isolated from its neighbors over a wide range of $t/J_{\rm H}$.
In particular, they stay nearly flat, with a shape that depends very weakly on the coupling. These two block-edge bands both carry $|C|=1$ and are especially robust.

Figure~\ref{fig2}(a-d) traces these two branches from weak to strong coupling, and Fig.~\ref{fig2}(e) plots their flatness, defined as the ratio of the upper adjacent gap to the bandwidth.
In what follows we refer to a band as flat when its flatness is $\gtrsim 10$. The lower direct gap is always finite but can be small, as enlarged in the inset of Fig.~\ref{fig2}(a). We therefore refer to these bands as isolated rather than well isolated. The two branches exhibit distinctly different evolutions with $t/J_{\rm H}$. As $t/J_{\rm H}$ is reduced, the 198th band evolves from a spectrally mixed weak-coupling regime at $t/J_{\rm H}=1$ [Fig.~\ref{fig2}(a)] into an isolated and practically flat $C=-1$ branch by $t/J_{\rm H}=0.39$ [Fig.~\ref{fig2}(b)], where it also shows a much more homogeneous Berry-curvature distribution. Its flatness sharpens dramatically near $t/J_{\rm H}\approx0.32$ [Fig.~\ref{fig2}(c,e)], where the branch switches from $C=-1$ to $C=+1$. This Chern number then persists toward the strong-coupling limit down to $t/J_{\rm H}=0.01$ [Fig.~\ref{fig2}(d)]. The 398th band, in contrast, remains isolated and flat across the entire scanned window with a stable Chern number $C=-1$.

Figure~\ref{fig2}(f-h) illustrates a further application of the same architecture. In the representative texture, the 7th band is perfectly flat with $E=-J_{\rm H}-3t$ in the strong-coupling limit but is not isolated there. As $t/J_{\rm H}$ increases, it becomes isolated. However, Fig.~\ref{fig2}(f) shows that it is still topologically trivial at $t/J_{\rm H}=0.43$. Near $t/J_{\rm H}=0.72$ [Fig.~\ref{fig2}(g)] it undergoes a narrow avoided crossing with the 6th band near $K$ and $K'$, with a finite direct gap [inset of Fig.~\ref{fig2}(g)], and acquires $C=-2$ that persists up to $t/J_{\rm H}=1$ [Fig.~\ref{fig2}(h)].

\textit{Real-space mechanisms of flatness.---}
Having established in Fig.~\ref{fig2} that the DHSKX supports robust $|C|=1$ branches, we now turn to the real-space wave functions. The overarching picture that emerges is a separation of roles: the perfectly flat $C=0$ precursors are core-localized states, whereas the recurring $|C|=1$ branches are governed by an extended phase organization. Figure~\ref{fig3} shows the two real-space signatures needed for this distinction. For clarity, only the dominant global-spin component is shown here.

For the trivial 7th band at $t/J_{\rm H}=0.43$ [Fig.~\ref{fig3}(a)], the wave function is concentrated near the center of the representative DHSKX unit cell and exhibits $\pi$-phase alternation on neighboring sites. This local antiphase pattern suppresses the net hopping amplitude out of the core region through destructive interference~\cite{PhysRevB.78.125104}, producing a perfectly flat $C=0$ band---the DHSKX analogue of compact localized states in frustrated-hopping flat bands. Unlike conventional compact localized states, however, this core orbital can become isolated under double exchange rather than remaining pinned to a band touching.

A qualitatively different and more robust organization underlies the isolated topological branches. In the 398th band at $t/J_{\rm H}=0.01$ [Fig.~\ref{fig3}(b)], and similarly in the 198th band at representative couplings, there is no strict nearest-neighbor sign alternation. Instead, the dominant weight organizes into three extended regions of nearly uniform phase, with relative offsets close to $\pm2\pi/3$---the threefold pattern expected from the magnetic $C_3$ organization of the triple-$Q$ texture. We refer to this pattern as \textit{phase clustering}. The phase diagnostic is quoted in the physical atomic-orbital basis of Eq.~\eqref{HDE}. 
Its origin is architectural. The $\pi$-locked opposite helicities make the in-plane spins of neighboring sites nearly antiparallel on a ring around each skyrmion core, 
suppressing precisely the bonds that would carry a core-centered circulation; the expelled winding is forced onto the surrounding network, where the magnetic $C_3$ pins it 
into the three $\pm2\pi/3$ clusters. Two controlled comparisons isolate this ingredient: a helicity interpolation that removes the sublattice helicity difference destroys the 
clustering and the strong flatness together while leaving the skyrmion cores, the $S^c$ landscape, and both topological charges intact, and analytic same-helicity triple-$Q$ 
textures---ordinary Bloch skyrmion crystals with the identical magnetic $C_3$---keep the winding confined to the cores and develop neither clustering nor comparable 
flatness.

Bond-resolved currents make the flattening mechanism explicit: the flat $|C|=1$ states carry circulating kinetic currents comparable to those of a dispersive band, yet their net transport cancels to one part in $10^{3}$ across the Brillouin zone, and the cancellation degrades in step with the helicity interpolation. Unlike compact localized states, whose flatness arises from local cancellation, these states are extended: the cluster network cancels net transport by distributed destructive interference while its circulating complex hoppings carry the Berry curvature. This organization is a band-wide property rather than a $\Gamma$-point feature: the fraction of kinetic bond weight connecting different phase domains remains low across the Brillouin zone for the flat $|C|=1$ states, clearly separated from dispersive reference bands and from spatially scrambled domain labels; the same organization persists for the 198th and 398th branches across representative couplings and survives in the reduced-period $6\times6$ analogue. Phase clustering is thus both forced and operative: the DHSKX creates and pins the clustered winding, and the clustering flattens the band.

\textit{Quantum geometry.---} Quantum geometry measures how closely a flat Chern band approaches the Landau-level limit~\cite{Roy2014,WangLiu2021}. For an isolated band, the quantum metric $g_{ab}(\mathbf{q})$ and the Berry curvature $\Omega(\mathbf{q})$ are the real and imaginary parts of the quantum geometric tensor~\cite{ProvostVallee1980,ParameswaranRoySondhi2013}, respectively. An ideal Chern band saturates the trace condition $\mathrm{tr}\,g(\mathbf{q})=|\Omega(\mathbf{q})|$ pointwise in the Brillouin zone~\cite{Roy2014}; deviations are quantified by
\begin{eqnarray}
\sigma_{\rm QG}=\frac{\int_{\rm BZ}[\mathrm{tr}\,g(\mathbf{q})-|\Omega(\mathbf{q})|]\,d^2q}{\int_{\rm BZ}|\Omega(\mathbf{q})|\,d^2q},
\label{eq_sigmaQG}
\end{eqnarray}
with $\sigma_{\rm QG}=0$ in the ideal limit.

We compute the quantum-geometric tensor on each band's isolated window, which includes $0.01\le t/J_{\rm H}\le 1$ for the 398th, 
$0.01\le t/J_{\rm H}\lesssim 0.39$ for the 198th, and $t/J_{\rm H}\gtrsim 0.43$ for the 7th after it isolates. The three branches separate sharply, and the contrast between the two block-edge bands is the central one. For the 398th, $\sigma_{\rm QG}$ is smallest at the strong-coupling edge and worsens steadily as $t/J_{\rm H}$ grows. Thus, despite remaining flat and isolated throughout, its quantum geometry is anchored in the adiabatic regime. This branch therefore plays the role of a broad adiabatic reference. The 198th reverses the trend. Its $\sigma_{\rm QG}$ minimum lies on the weaker-coupling side of the $C$-sector switch at $t/J_{\rm H}\approx 0.32$ and, at $t/J_{\rm H}=0.39$, coincides with the band crossing the flatness threshold and reaching its most homogeneous Berry-curvature distribution, identifying the post-transition branch as a beyond-adiabatic intermediate-coupling regime. Quantitatively, at $t/J_{\rm H}=0.39$ the 198th band achieves $\sigma_{\rm QG}\approx0.122$, compared with $\sigma_{\rm QG}\approx1.86$ at $t/J_{\rm H}=0.01$ for the 398th adiabatic reference, directly confirming the fifteen-fold improvement in quantum geometry. This point serves as the representative single-particle reference throughout. The 7th band is yet different. Its trivial segment has nearly zero average Berry curvature and poor normalized geometry, and only the avoided approach to the 6th band near $t/J_{\rm H}\approx 0.72$ produces a narrow low-$\sigma_{\rm QG}$ pocket coincident with the $C=-2$ window.

This beyond-adiabatic characterization is also visible directly in the spin content of the wave functions. Projecting the fixed-global-basis spinors onto the local spin frame with $P_i^\pm=(1\pm\mathbf{S}_i\cdot\boldsymbol{\sigma})/2$, we find that the 198th band at $t/J_{\rm H}=0.39$ carries a local nonadiabatic weight of $0.219$ in the spin sector opposite to the lower-block adiabatic locking, with a global-basis subdominant spin weight of $0.263$. 
The former is computed via $P_i^+$ in the texture-corotating local frame and the latter measures the minority-component weight in the fixed global spin basis. They differ because the local quantization axis rotates across the unit cell. In the post-transition interval $0.32\le t/J_{\rm H}\le0.39$, this spin mixing grows while $\sigma_{\rm QG}$ and the normalized Berry-curvature nonuniformity fall from $2.31$ to $0.122$ and from $6.31$ to $0.653$, respectively. We therefore use ``beyond adiabatic'' operationally for this geometry-improved, phase-clustered regime rather than for a strong-coupling spin-following limit.

Band-projected exact diagonalization at $\nu=1/3$ provides consistent finite-size evidence for Laughlin-type FCI physics~\cite{TangMeiWen2011,WangGuGongSheng2011,Ghorashi2023}, 
and ties it to the quantum geometry: across a coupling scan the Laughlin signatures open only on the geometry-optimized branch, in the window $0.37\lesssim t/J_{\rm H}\lesssim0.44$ 
around the $\sigma_{\rm QG}$ minimum, and are absent on the pre-transition branch despite its far larger flatness ratio. 
Across accessible non-anomalous tori up to $N_\phi=30$ ($10\times3$, $N=10$; momentum-sector dimension $1.0\times10^{6}$), the three lowest states occupy the 
Laughlin momentum sectors; along the clean $N_y=3$ sequence, the gap-to-width ratio $\Delta_{34}/W_3$ grows from $2.00$ to $2.31$ and 
saturates within the largest sizes studied. The three-state manifold stays gapped over the entire boundary-twist torus (minimum $\Delta_{34}=0.033$ on $6\times3$ and $0.020$ on $8\times3$), reliable many-body Chern diagnostics give $C_{\rm MB}=+1$ on the clean $6\times2$ torus and on the sector-tracked $6\times3$ torus, and two independent countings match the $\nu=1/3$ generalized Pauli principle exactly: the particle entanglement spectrum~\cite{LiHaldane2008} in a narrow window slightly below $t/J_{\rm H}=0.39$, and the $N=5$ quasihole spectrum at $N_\phi=18$ ($126$ states, $7$ per momentum sector). These diagnostics are insensitive to gate screening and to the residual band dispersion. A definitive thermodynamic identification remains beyond the present finite sizes.

\textit{Conclusion and outlook.---}
We have shown that a double-helix skyrmion crystal under double exchange provides a discrete, architecture-based route to flat Chern bands. The central result is the phase-clustered $|C|=1$ mechanism with its two layers---the $\pi$-helicity locking expels the winding from the cores, and the magnetic $C_3$ symmetry pins the threefold clustered winding, which flattens the band by 
distributed destructive interference---realized most prominently in the lower block-edge (198th) branch. This branch reaches its best quantum geometry at intermediate coupling, surpassing the adiabatic strong-coupling reference and identifying the DHSKX as a beyond-adiabatic route to flat Chern bands. In the geometry-optimal window, band-projected exact diagonalization gives finite-size evidence consistent with $\nu=1/3$ Laughlin-type FCI physics. The same phase-clustered organization also appears in the 398th branch, and survives in the $6\times6$ reduced-period analogue (70th and 142nd bands), showing that it is a property of the DHSKX architecture.
Perfectly flat $C=0$ precursors are core-localized antiphase orbitals, and the $C=-2$ window is a secondary feature of the same representative texture. The relevant structure is the DHSKX architecture itself.

Because the double-exchange problem on a fixed DHSKX background reduces to a tight-binding network with site-dependent complex hoppings and exchange fields, the architecture can be reproduced without a self-organized magnetic ground state. As a minimal concrete example, the confined $6\times6$ DHSKX analogue maps onto a 72-node honeycomb topolectric circuit~\cite{Lee2018,Imhof2018,Wu2022} whose complex admittances $Y_{ij}\propto\cos(\theta_{ij}/2)\,e^{iA_{ij}}$ are set by the local spin canting angle $\theta_{ij}$ and Berry-connection phase $A_{ij}$ of the DHSKX texture, and are readily synthesized from capacitor-INIC (negative-impedance converter) building blocks; acoustic resonator arrays and photonic lattices offer analogous routes. More broadly, discrete noncoplanar textures with sublattice-resolved topological charges constitute a fruitful class of architectures for flat Chern band engineering, complementary to moir\'e and continuum-skyrmion approaches.

\textit{Acknowledgements.---}
We thank Y.-F. Wang for helpful discussion. This work is supported by the National Key R$\&$D Program of China (Grants No. 2022YFA1402704), by the National Natural Science Foundation of China (Grants Nos. 12274187, 12304176, 12247101), by the Shenzhen Fundamental Research Program (Grant Nos. JCYJ20220818100405013 and JCYJ20230807093204010) and by the Natural Science Foundation of Jiangsu Province (Grant No. BK20241929).

\section*{Data Availability}
The data that support the findings of this work are available from the corresponding author upon reasonable request.


%

\end{document}